\documentstyle[prl,preprint,multicol,aps]{revtex} \input epsf

\begin{document}

\draft 

\title{Onset Temperature of Slow Dynamics in Glass Forming Liquids}

\author{Srikanth Sastry$^{*}$}
\address{Jawaharlal Nehru Centre for Advanced Scientific Research, 
Jakkur Campus, Bangalore 560064, INDIA}                            

\maketitle

\begin{abstract}
The behaviour of a model glass forming liquid is analyzed for a range
of densities, with a focus on the temperature interval where the
liquid begins to display non-Arrhenius temperature dependence of
relaxation times. Analyzing the dynamics along with properties of
local potential energy minima sampled by the liquid, a crossover or
onset temperature $T_s$ is identified below which the liquid manifests
{\it slow dynamics}, and a change in the character of typical local
potential energy minima.
\end{abstract} 




The phenomenology of glass formation, whereby a liquid cooled to low
temperatures manifests increasingly sluggish dynamics, and at low
enough temperature transforms into an amorphous solid, has come under
increasing study in recent years\cite{reviews,trieste}. In addition to
the glass transition temperature $T_g$ where the liquid transforms to
a glass in the laboratory, much attention has been devoted to the
Kauzmann or {\it ideal} glass transition temperature
$T_K$\cite{dimarz,adam-gibbs,kirk,fs,heuer,parisi,scala},
which locates a putative thermodynamic phase transition underlying
the observed glass transition, and the mode coupling
critical temperature $T_c$\cite{mct,kob,fsonMCT}, which is now
commonly viewed as marking the crossover to a regime of `activated
dynamics' discussed long ago by
Goldstein\cite{Goldstein69}. Relatively less attention has been paid
to the range of temperatures (or equivalently, a crossover
temperature) where the nature of the dynamics changes from what may be
termed as `normal' to what is commonly referred to as `slow
dynamics'\cite{sastrynature,fld,jonsson}. The purpose of this letter is to
demonstrate, on the basis of computer simultions of a model liquid,
that such a crossover temperature can be identified by various
indicators that mark a change in the nature of local potential energy
minima the liquid samples, in addition to changes in the dynamics
itself.

Results presented here are from molecular dynamics simulations of a
model liquid which is a binary mixture consisting of $204$ type $A$
and $52$ type $B$ particles. The particles interact {\it via} the
Lennard-Jones (LJ) potential, with parameters
$\epsilon_{AB}/\epsilon_{AA} = 1.5$, $\epsilon_{BB}/\epsilon_{AA} =
0.5$, $\sigma_{AB}/\sigma_{AA} = 0.8$, and $\sigma_{BB}/\sigma_{AA} =
0.88$, and $m_B/m_A = 1$.  This system has been extensively studied as
a model glass
former\cite{kob,kobhet,sastrynature,fs,parisi,sastryprl}. The LJ
potential is modified with a quadratic cutoff and shifting at
$r_c^{\alpha \beta} = 2.5 \sigma_{\alpha \beta}$\cite{stoddard}.  All
quantities are reported in reduced units, length in units of
$\sigma_{AA}$, volume $V$ in units of $\sigma_{AA}^{3}$ (density $\rho
\equiv N/V$, where $N$ is the number of particles, in units of
$\sigma_{AA}^{-3} \equiv \rho_0$), temperature in units of
$\epsilon_{AA}/k_B$, energy in units of $\epsilon_{AA}$ and time in
units of $\tau_m \equiv (\sigma_{AA}^2 m/\epsilon_{AA})^{1/2}$, where
$m = m_A = m_B$ is the mass of the particles. Molecular dynamics
simulations are performed over a wide range of temperatures at each
density, and local energy minimizations are performed for $1000$ ($k_B
T/\epsilon_{AA} < 1.$) or $100$ ($k_B T/\epsilon_{AA} > 1.$)
configurations to obtain typical local energy minima or `inherent
structures'\cite{inh} sampled by the liquid. Inherent structures are
obtained for a subset of temperatures at intervals that vary between
$0.2$ to $0.4$ near the crossover $T$, which thus limit the precision
of estimation, in addition to the inherent lack of sharp definition of
a crossover temperature. Further details concerning the simulations
may be found in \cite{sastryprl}, where a study of the dynamics and
thermodynamics of this model was reported over a wide range of
densities and temperatures.

Data from these simulations are analyzed here for five densities to
study the crossover temperature where the liquid begins to manifest
slow dynamics. The dynamics is probed by calculating the self
intermediate scattering function $F_s(k,t)$ for the $A$ particles,
which is the Fourier transform at wavevector $k$ of the van Hove self
correlation function
\begin{equation} \label{eq:ddcf}
G_s(r,t) = {1 \over N_A} \sum_{i = 1}^{N_A} \left<\delta(|{\bf r}_i(t)-{\bf r}(0)| - r)\right>, 
\end{equation}
which describes the probability of finding at time $t$ a particle at
distance $r$ from its location at $t = 0$. At each density, the
wavevector studied is close to the first peak of the structure
factor. Relaxation times $\tau(T)$ are obtained as a function of
temperature $T$, at each density, from $F_s(k,t)$ with the definition,
$F_s(k,\tau) = 1/e$. At high $T$, $\tau$ display Arrhenius behavior
({\it i. e.}, $\tau = \tau_0 \exp(E_0/k_B T)$ where $\tau_0$ and $E_0$
are constants) as shown in Fig. 1(a). Deviations from Arrhenius
behavior can be seen clearly by defining a $T$-dependent `activation energy'
\begin{equation}
E(T) = k_B T\log(\tau/\tau_0)
\end{equation}
which has the constant value $E_0$ for $T$ where the Arrhenius form is
valid. High $T$ values of $\tau$ are used to estimate $\tau_0$ and
$E_0$ which are then used to transform the full data set, following
the procedure in \cite{sastrynature}. The resulting $E(T)$ data
(normalized to $E_0$) are shown in in Fig. 1(b), which display for all
densities $\rho$ the crossover to non-Arrhenius behavior, similar to
the finding in \cite{sastrynature} for $\rho = 1.2 \rho_0$. The crossover $T$
identified this way are shown in Fig. 2. Also shown are the ideal
glass transition line (`$T_K$'; from \cite{sastryprl}) and the VFT
temperature $T_0$ obtained from diffusivity data in \cite{sastryprl}.
The mode coupling temperatures $T_c$ have not been estimated as a
function of density in this study, but will be presented in a future
article. At $\rho = 1.2 \rho_0$, the value of $T_c$ estimated in \cite{kob}
is $k_B T_c/\epsilon_{AA} = 0.435$, while from the present study, $k_B T_s/\epsilon_{AA} = 1.0$ and $k_B T_K/\epsilon_{AA} =
0.2976$. Therefore, the mode coupling $T_c$ at this density is less than
half the onset temperature $T_s$, and higher than the ideal glass transition 
temperature $T_K$ by roughly $50 \%$.

Properties of potential energy minima sampled by the liquid show a
corresponding crossover behavior, which is now discussed. A number of
recent studies\cite{speedy,fs,heuer,parisi,ruocco,crisanti,scala,sastryprl,sastrytrieste}
have focussed on analyzing the thermodynamics of glass forming liquids
on the basis of a decomposition of the configurational space of the
liquid into {\it basins}, as a route to estimating the configurational
entropy, which is given by the multiplicity of such basins. In the
inherent structure approach\cite{inh}, basins are identified as basins
of the local potential energy minima (inherent structures), which
permits the partition function to be written as (details may be found
in \cite{fs,heuer,sastryprl})
\begin{equation}\label{eq:pfn} 
Q_N(\rho,T) = \int d\phi \Omega(\phi) ~exp\left(-\beta N (\phi + f_{basin}(\phi,T))\right),
\end{equation} 
where $\phi$ is the potential energy per particle of the inherent
structure, $\Omega(\phi) d\phi$ is the number of inherent structures
in the range ($\phi, \phi + d\phi$), $\exp(-\beta~f_{basin}(\phi,T))$
is given by the restricted partition function integral over the basin
of an inherent structure of energy $\phi$. The configurational entropy
density is defined as ${\cal S}_c(\phi) \equiv {k_B \over N} \ln \Omega(\phi)$ and
is related to the $T$ dependent configurational entropy by $S_c(T) =
\int d\phi~ {\cal S}_c(\phi) P(\phi,T)$. The
probability of sampling an inherent structure of energy $\phi$ is given by 
\begin{equation} \label{eq:pofe}
P(\phi,T) = exp\left(-\beta N (\phi + f_{basin}-{\cal S}_c)\right)/Q_N(\rho,T).
\end{equation} 
The distribution of inherent structures sampled is calculated directly
from simulations, by performing local energy minimizations for a
sample of equilibrated configurations. In addition, the total free
energy of the system (and hence $Q_N$) is calculated from
thermodynamic integration as outlined in \cite{sastryprl}. Knowledge
of the basin free energy $f_{basin}$ is then sufficient to invert the
above relationship to obtain the configurational entropy density
${\cal S}_c$. One must obtain the same estimates of ${\cal S}_c$ from
$P(\phi,T)$ at different temperatures, in the range of $\phi$ values
where the corresponding $P(\phi,T)$ overlap. In previous
work\cite{fs,heuer,sastrytrieste} it has been observed that one indeed
obtains overlapping ${\cal S}_c$ estimates from different temperatures
at sufficiently low temperatures. Results in these analyses was based
on evaluating the basin free energy with a harmonic approximation,
wherein the basin free energy is given in terms of the frequencies
$\nu_i$ which are related to the eigenvalues $\lambda_i$ of the Hessian
matrix (matrix of second derivatives of the potential, defined at the
energy minima) by $2 \pi \nu_i = \sqrt{\lambda_i}$. The basin free energy 
is given in this approximation by:
\begin{equation}\label{eq:fbasin}
\beta f_{basin}  = {1 \over N} \sum_i^{3N} \ln {h \nu_i \over k_B T} 
\end{equation}
The eigenvalues $\lambda_i$ are obtained numerically for the sample of
inherent structures obtained at each temperature and density.
Estimates of ${\cal S}_c$ from different temperatures for each density
are shown in Fig. 3. It is seen that at each density, the ${\cal S}_c$
estimates from low temperatures overlap with each other, but estimates
from higher temperatures fail to do so (each $T$ is not, however,
individually labeled in the figure). The harmonic approximation for
describing the properties of basins therefore fails at high
temperatures, indicating a qualitative change in the nature of the
basins sampled at temperatures below and above a crossover
temperature. The crossover temperatures estimated in this fashion are
shown in Fig. 2. The criterion used here, as well as in Fig. 4, 5, is
stated in the figure caption.

It has been observed\cite{heuer,speedyfr,sastrytrieste} that
distribution of inherent structures $\Omega(\phi)$ is well described
by a Gaussian (equivalently, the configurational entropy density
${\cal S}_c$ by an inverted parabola). Figure 3 shows fits to ${\cal
S}_c$ at each density of the form,
\begin{equation}\label{eq:scdos} 
{\cal S}_c(\phi)/k_B = \alpha - {(\phi - \phi_0)^2\over \sigma^2}.
\end{equation} 
The parameters $\alpha$, $\phi_0$ and $\sigma$ depend on the density
$\rho$ but not on the temperature $T$.  Based on the assumption of a
Gaussian $\Omega(\phi)$ and a constant (with respect to $\phi$) $\beta
f_{basin}$ in Eq. (\ref{eq:pfn}) it has been
observed\cite{speedyfr,heuer} that the average inherent structure
energy $\phi(T)$ should vary with temperature as $\phi(T) = \phi_0 -
\sigma^2/2 k_B T$. However, the $\phi$ dependent component of $\beta
f_{basin}$ (arising from the dependence of the frequencies $\nu_i$ on
the basin occupied, indexed by its energy) is not constant but varies
roughly linearly with $\phi$\cite{sastrynature2} (see also
\cite{fsaging,water}). Hence, the form
\begin{equation} \label{eq:fvib}
\beta f_{basin}(\phi) = \beta f_{0} + \delta f (\phi - \phi_0),
\end{equation}  
where $\beta f_0$ has no $\phi$ dependence, is employed to describe
the $T$ and $\phi$ dependence of $\beta f_{basin}$. Inclusion of this
$\phi$ dependence does not alter the prediction of the $1/T$
dependence but redefines the quantitative expression to
\begin{equation} \label{eq:phi}
\phi (T) = \phi^{eff}_0 - \sigma^2/2 k_B T,
\end{equation} 
where the {\it effective} mean IS energy $\phi^{eff}_0 = \phi_0 -
\delta f \sigma^2/2$. It has been shown elsewhere\cite{sastrynature2}
that Eq. (\ref{eq:phi}) indeed describes well the average inherent
structure energies, at low temperatures, calculated directly from
simulations. However, at high temperatures, one observes deviations
from the $1/T$ behavior (see also \cite{sastrytrieste}). A {\it scaled
plot} of $\phi (T) - \phi^{eff}_0$ {\it vs.} $\sigma^2/2 k_B T$ is
shown in Fig. 4. In order to identify the crossover temperature as
accurately as possible, the fit parameters involved, $\alpha$,
$\phi_0$, $\sigma$ and $\delta f$ have been estimated by fitting
jointly the low temperature data sets of ${\cal S}_c(\phi)$
(Eq.(\ref{eq:scdos})), $\beta f_{basin}$ (Eq.(\ref{eq:fvib})) and $\phi
(T)$ (Eq.(\ref{eq:phi})). The fit parameters are shown in Table
I. Crossover temperatures where $\phi(T)$ exhibits deviations from the
$1/T$ form are shown in Fig. 2. The analysis here makes more precise,
and quantitative, the observation in \cite{sastrynature} of a
correlation between the onset of non-Arrhenius relaxation and a change
from nearly constant to $T$ dependent values of the average inherent
structure energies.

When inherent structure basins sampled by the liquid are well
described by a harmonic approximation, one must expect that the
potential energy per particle $u(T)$ of the equilibrated liquid is
related to the inherent structure energy $\phi(T)$ by $u(T) = \phi(T)
+ 3/2 k_B T$. Thus significant differences between the quantity
$\Delta u(T) = u(T) - 3/2 k_B T$ and $\phi(T)$ also indicate
anharmonicity of the basins sampled\cite{andrea,sastrytrieste}.
Fig. 5 shows $\Delta u(T)$ plotted against $\phi(T)$ for each of the
studied densities. The range of inherent structure energies sampled is
non-monotonic in density, with the lowest energies obtained at $\rho =
1.2 \rho_0$.  For each density, there is a low energy range for which the
harmonic expectation (indicated by the solid line) is met to a good
extent (the deviations from the harmonic expectation in this range is
less than $4 \%$ in all cases). At higher energies, it is seen that a
fairly sharply defined energy (or temperature) exists where the two
quantities deviate significantly. Such deviation permits the
identification of a crossover temperature. The crossover temperatures
so obtained are plotted in Fig. 2.

Comparison of the various crossover temperatures in Fig. 2 identified
from characterizing the dynamics (Fig. 1) and the inherent structures
sampled (Fig.s 3, 4, 5) show that the crossover from `normal' to
`slow' dynamics is associated with a change in the nature of local
potential energy minima the liquid samples. The crossover is
particularly broad for $\rho = 1.35$ which correspondingly shows the
poorest agreement between different estimates. The correspondence of
changes in IS properties with liquid {\it structure} was studied in
\cite{jonsson}, and more recently in \cite{water} for water. Some
analysis of this crossover has been made for spinglass models, {\it
e. g.} \cite{crisanti,sharontrieste}. In \cite{crisanti} it is shown
that finite range spin glass models show behavior of average energies
of local minima which are similar to liquids. Similar findings for the
nearest neighbor $\pm J$ spinglass model have been reported in
\cite{sharontrieste}. Analysis of the {\it free energy} landscape for
the hard sphere fluid\cite{chandantrieste} identify a crossover
density where glassy free energy minima come into existence, which may
be related to the crossover discussed here. Frustration limited domain
theory\cite{fld} discusses an avoided critical point temperature, which 
has been identified with the crossover temperature discussed here. 

Nevertheless, the precise nature of the crossover is at present
unclear. Although Fig.s 3--5 indicate a change from `harmonic' basins
at low temperature to those that are anharmonic at $T$ above the
crossover $T$, detailed analyses\cite{schroeder,andrea,saddles2}
indicate that at temperatures between mode coupling $T_c$ and the
crossover temperature $T_s$, the dynamics appears {\it via} motion
along directions in configuration space with negligible energy
barriers, the barriers being instead {\it
entropic}\cite{chandantrieste}. Therefore, even though the dynamics in
the range of $T_c$ to $T_s$ is `landscape
influenced'\cite{sastrynature} it is `landscape dominated' only below
$T_c$\cite{Goldstein69}. Notwithstanding the quantitative measures
presented here, the precise sense in which the harmonicity of local
minima changes across $T_s$ is unclear. Very little theoretical effort
has been directed at understanding the nature of the crossover at
$T_s$. Results presented here, along with related evidence cited, call
for a focus on phenomena pertaining to the onset temperature of
slow dynamics, in addition to more thoroughly studied phenomena at 
lower temperatures. 

\centerline{\Large
\begin{tabular}{|c||c|c|c|c|} \hline
\large{$\rho/\rho_0$} & \large{$\alpha$} & \large{$\phi_0/\epsilon_{AA}$} & \large{$\sigma/\epsilon_{AA}$} & \large{$\delta f \epsilon_{AA}$} \\ 
\hline 
\hline 
    1.10 &  1.020 &  -6.579 &  0.312 &  0.616 \\ \hline
    1.15 &  0.963 &  -6.680 &  0.379 &  0.603 \\ \hline
    1.20 &  0.921 &  -6.700 &  0.470 &  0.455 \\ \hline
    1.25 &  0.875 &  -6.642 &  0.550 &  0.350 \\ \hline
    1.35 &  0.860 &  -6.080 &  0.870 &  0.159 \\ \hline
\end{tabular}
}

\bigskip

\noindent TABLE I: Fit parameters quantifying the distribution of
inherent structures, and the dependence of vibrational free energy on
the inherent structure energy, described in Eq.s (Eq.(\ref{eq:scdos}))
and (Eq.(\ref{eq:fvib})).

\begin{figure}[b]
\hbox to\hsize{\epsfxsize=1.0\hsize\hfil\epsfbox{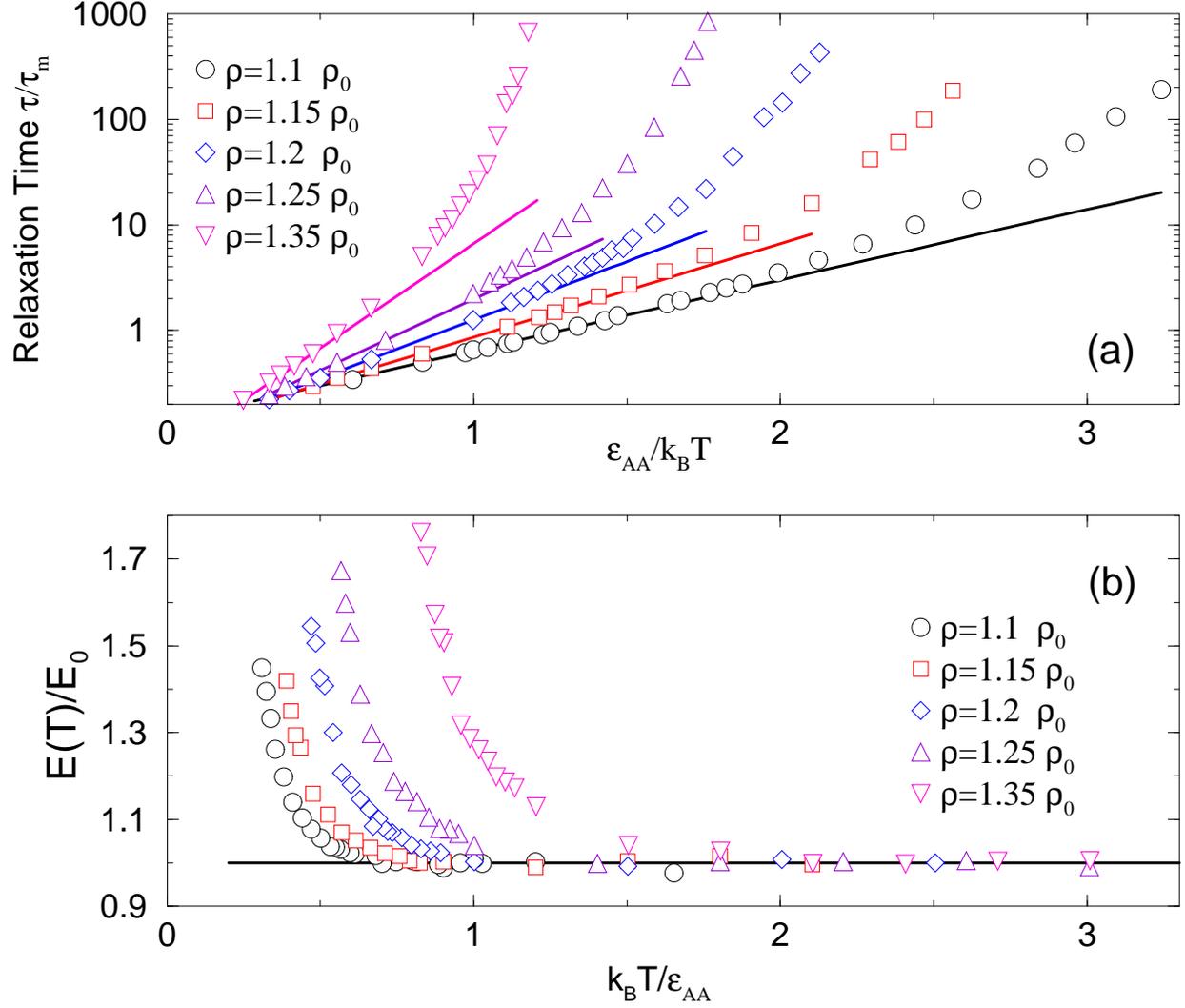}\hfil}
\caption{\narrowtext (a) Arrhenius plot of relaxation times for
different densities. At each density, a crossover from Arrhenius
behavior at high temperature to super-Arrhenius behavior is observed. (b)
Temperature dependent `activation energies' $E(T)$ normalized to the 
high temperature value $E_0$.}
\label{fig1}
\end{figure}

\begin{figure}[b]
\hbox to\hsize{\epsfxsize=1.0\hsize\hfil\epsfbox{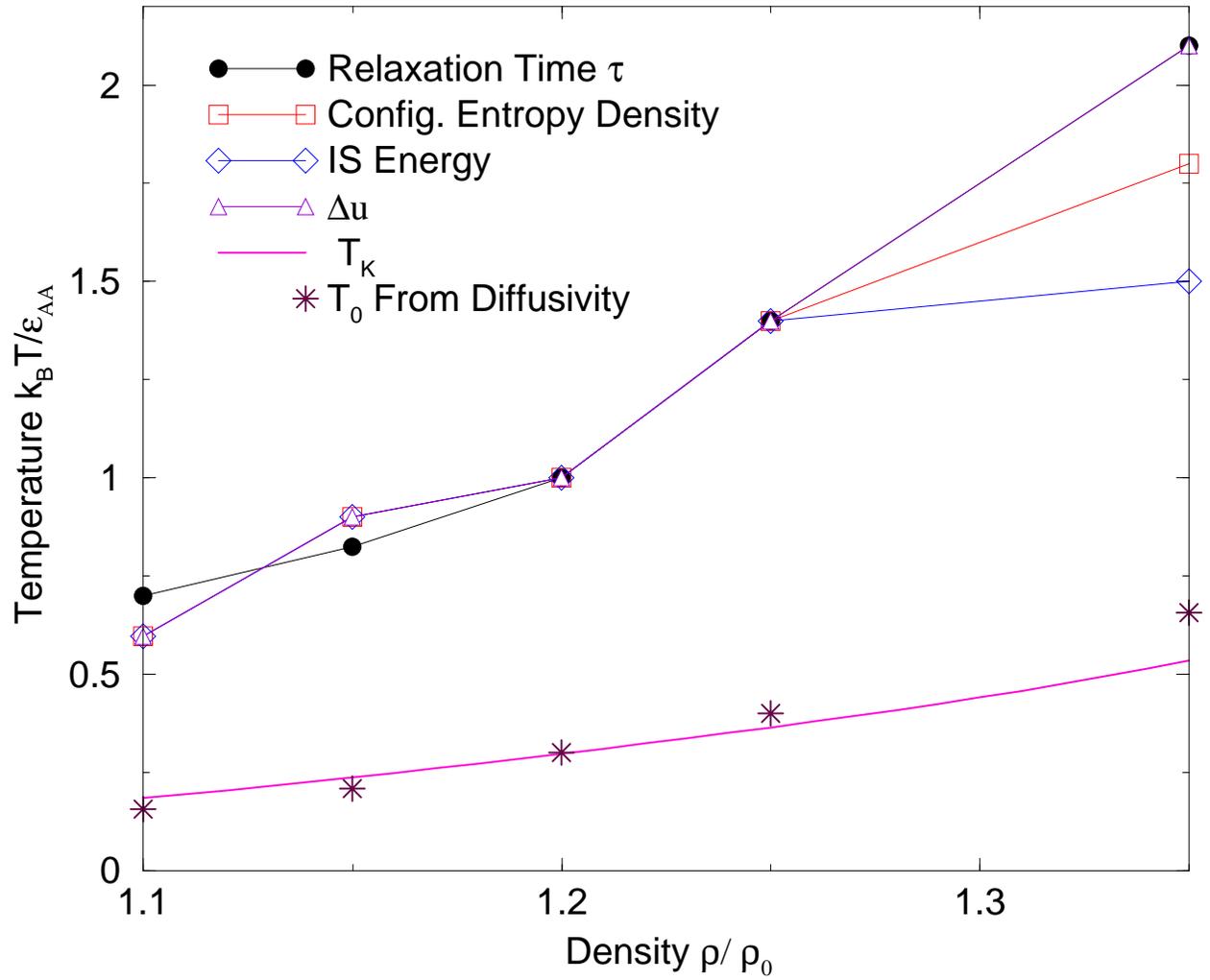}\hfil}
\caption{\narrowtext Crossover temperatures estimated from (i)
relaxation times(Fig. 1), (ii) Configurational entropy density ${\cal
S}_c(\phi)$ (Fig. 3), (iii) $T$ dependence of IS energies (Fig. 4) and
(iv) the energy difference between instantaneous and IS configurations
(Fig. 5).}
\label{fig2}
\end{figure}

\begin{figure}[b]
\hbox to\hsize{\epsfxsize=1.0\hsize\hfil\epsfbox{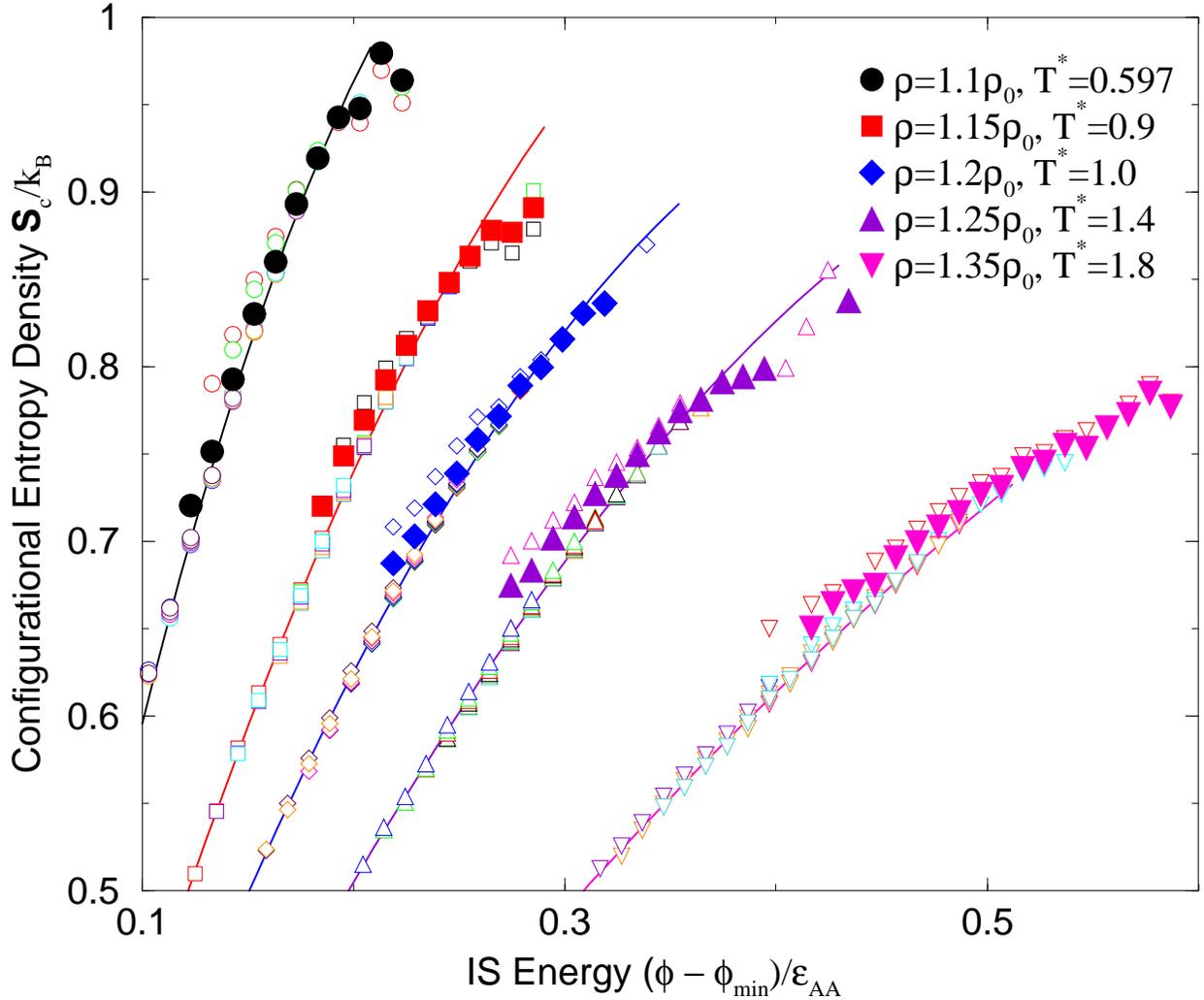}\hfil}
\caption{\narrowtext Configurational entropy density ${\cal
S}_c(\phi)$ estimated as described in the text. $\phi_{min}$ is the
(lower) value of $\phi$ where ${\cal S}_c(\phi) = 0$ from
Eq.(\ref{eq:scdos}). At each density (each overlapping set of curves),
${\cal S}_c(\phi)$ estimates obtained from a range of temperatures are
shown (not individually labeled). $T^* \equiv k_B T/\epsilon_{AA}$ in
the legends.  It is seen that at each density, a number of data sets
overlap with each other to a very good extent (these correspond to low
temperatures) while some data sets show much poorer overlap (these
correspond to high temperatures). The failure of high temperature
${\cal S}_c(\phi)$ curves to overlap with those at lower $T$ signals a
crossover in behavior. Data points for the lowest temperature where
such lack of overlap is discernible are displayed using filled symbols
and labeled. These temperatures provide estimates of the crossover
temperature.}
\label{fig3}
\end{figure}

\begin{figure}[b]
\hbox to\hsize{\epsfxsize=1.0\hsize\hfil\epsfbox{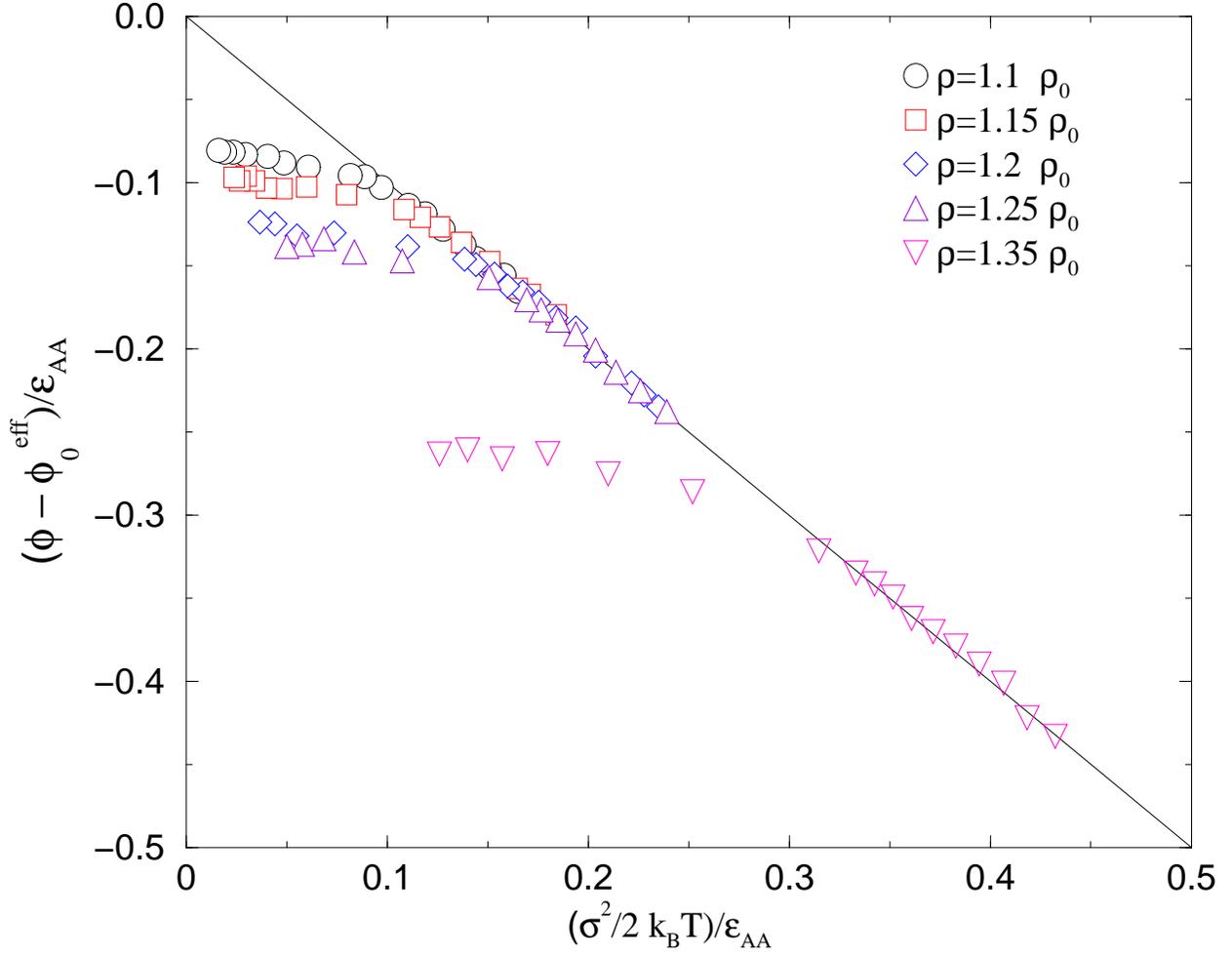}\hfil}
\caption{\narrowtext Scaled plot of the $T$ dependence of IS energies
described in the text. Deviations from the expected linear behavior at
high $T$ marks the crossover. The crossover temperature is
identified as the lowest $T$ for which the IS energies show
significant deviation from the low $T$ linear behavior.}
\label{fig4}
\end{figure}

\begin{figure}[b]
\hbox to\hsize{\epsfxsize=1.0\hsize\hfil\epsfbox{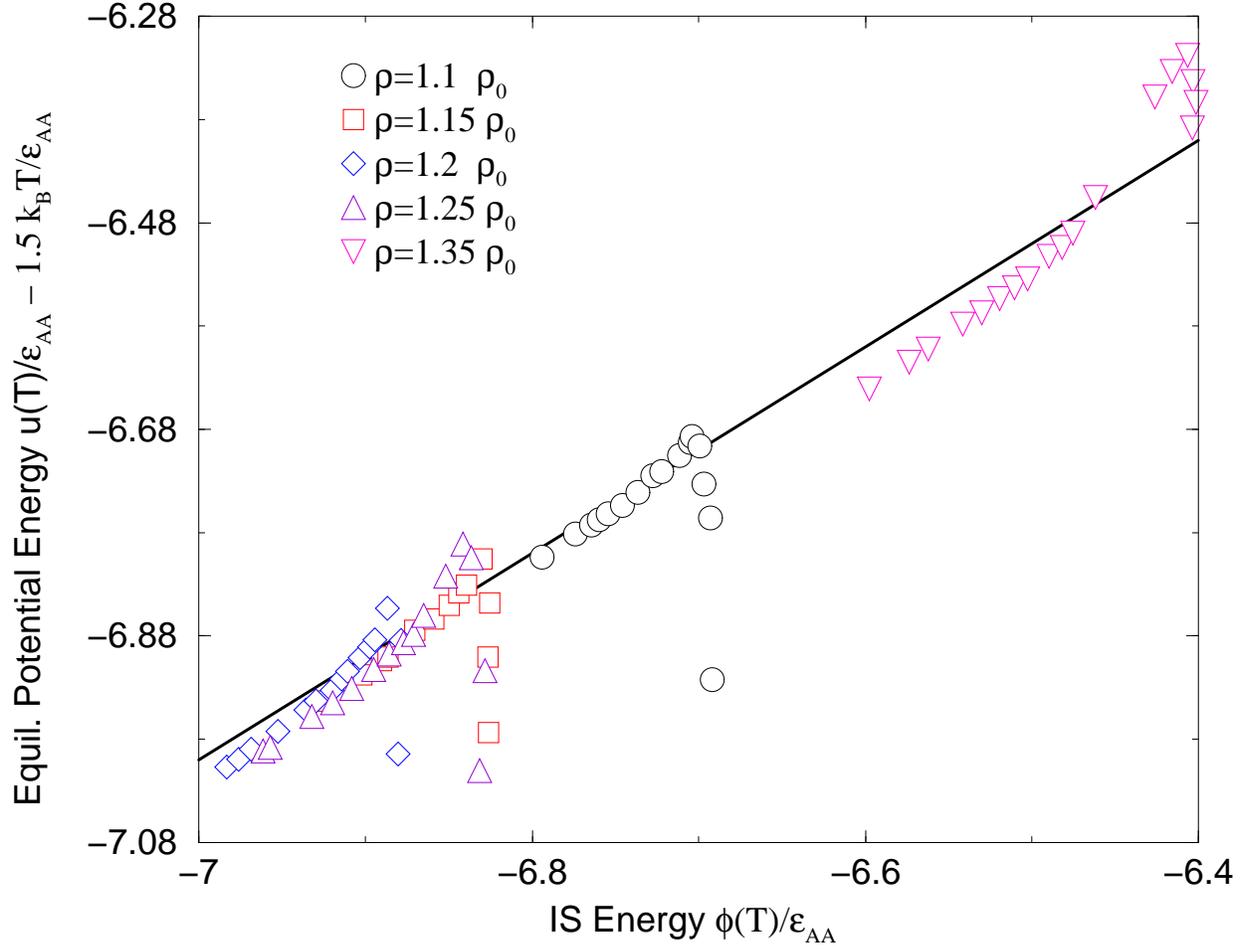}\hfil}
\caption{\narrowtext The equilibrium potential energy $u(T)$ minus $1.5 k_B
T$ (the harmonic expectation) is shown against IS energies. When the
harmonic approximation is valid one expects the data to fall on a
straight line of unit slope. At each density the temperature where 
$u(T) - 1.5 k_B T$ shows a maximum before dropping below the reference line
is identified as the crossover temperature.}
\label{fig5}
\end{figure}

\end{document}